# Superconducting properties of the oxygen-vacant iron oxyarsenide TbFeAsO$_{1-\delta}$ from underdoped to overdoped compositions

Y.G. Shi,[1,2,*] S. Yu,[3] A.A. Belik,[1,2] Y. Matsushita,[4] M. Tanaka,[4] Y. Katsuya,[5] K. Kobayashi,[4] Y. Hata,[6] H. Yasuoka,[6] K. Yamaura,[2,3] E. Takayama-Muromachi[1,2,3]

[1] International Center for Materials Nanoarchitectonics (MANA), National Institute for Materials Science, Tsukuba, Ibaraki 305-0044, Japan.
[2] JST, Transformative Research-Project on Iron Pnictides (TRIP), Tsukuba, Ibaraki 305-0044, Japan.
[3] Superconducting Materials Center, National Institute for Materials Science, 1-1 Namiki, Tsukuba, 305-0044 Ibaraki, Japan
[4] NIMS Beamline Station at SPring-8, National Institute for Materials Science, 1-1-1 Kouto, Sayo-cho, Sayo-gun, Hyogo 679-5148, Japan
[5] SPring-8 Service Co. Ltd., 1-1-1 Kouto, Sayo-cho, Sayo-gun, Hyogo 679-5148, Japan
[6] Department of Applied Physics, National Defense Academy, 1-10-20 Hashirimizu, Yokosuka 239-8686, Japan.

**Abstract**

A wide-range doping was achieved by a high-pressure method for TbFeAsO$_{1-\delta}$ from "under doped" to "over doped" superconducting compositions throughout the optimized superconductivity ($T_c$ of 44 K). $T_c$ vs. $\delta$ shows a dome-shaped feature, while $T_c$ vs. the lattice constant likely follows a unique empirical curve over the doping range. The relatively large amount of oxygen vacancies up to ~0.3 per the formula unit was introduced possibly because of the smaller replacement Tb than the other *Ln* (rare-earth element) in the *Ln*FeAsO$_{1-\delta}$ system.

**Introduction**

Discovery of high-$T_c$ superconductivity in the iron oxypnictide $Ln$FeAsO probably caused one of the highest impacts in materials science since 1986, the year copper oxide superconductor was discovered [1,2]. Reflecting the significant attention, large number of manuscripts has been submitted to the preprint archive about the superconductor since then [3]. It is also noticeable that development of varieties of the iron superconductor advanced swiftly; for example such as the 'infinite' system $A$Fe$_2$As$_2$ [4,5], the iron phosphide oxide [6,7], and the iron selenide [8,9] were discovered to be superconducting within several months right after the discovery. However, the earlier system $Ln$FeAsO still holds the highest $T_c$ record among them and keeps capturing scientific and practical attentions: the highest $T_c$ to date is 56 K for SmFeAs(O,F) [10]. Besides, the upper critical field is remarkably high [11], suggesting many possibilities of future applications. Meanwhile, we probably have a broad consensus that crystal growth of the doped $Ln$FeAsO is highly challenging. Due to lack of the high-quality crystals in experimental studies, a quality-improved and carefully controlled polycrystalline sample still holds high values for continuous progress of understandings of the iron superconductor.

The parent material $Ln$FeAsO crystallizes in a layered structure (ZrCuSiAs-type) [7], consisting of alternative stacking of Fe$_2$As$_2$ layer and $Ln_2$O$_2$ layer along a crystallographic direction. The Fe$_2$As$_2$ layer plays an essential role to establish the 3$d$ multi bands which are directly responsible for the superconductivity, and the insulating $Ln_2$O$_2$ layer plays a role as a charge-carriers source by accommodating doped elements and oxygen vacancies [12]. There are many studies in progress, however nature of the superconductivity seems to be under debate: superconducting symmetry was suggested to be fully gaped, while a pseudo-gap with line-node was suggested instead [13]. A delicate balance between the degree of antiferromagnetic 3$d$ spin fluctuations and the pair-breaking ferromagnetic interactions is possibly significant in determination of $T_c$ [12]. On the other side, superconducting properties are highly sensitive to local structure environments such as the coordination of iron by arsenic [14]. Although systematic studies are achieved in a wide variety of ways to some extent, further studies seems be needed to elucidate the principal physics of the superconductivity in the $Ln$FeAsO system.

In this study, we focused on TbFeAsO$_{1-\delta}$ in which oxygen vacancies are introduced to turn on the superconductivity without fluorine usually doped with. We found that $T_c$ goes up and then goes down in a systematic way with increasing the oxygen vacancies concentration. Indeed, a bell shaped feature in $T_c$ vs. $\delta$ was observed as well as what was found for the copper oxide superconductor. To our best knowledge, this would be the first observation of the bell shaped feature in the $Ln$FeAsO system.

**Experimental**

Details of the sample preparation for TbFeAsO$_{1-\delta}$ were reported elsewhere [15]. A high-pressure synthesis method was used. We prepared samples at $\delta$ = 0, 0.10, 0.15, 0.20, 0.25 (the first group, notated as #1), and several months later we synthesized again at $\delta$ = 0.10, 0.15, 0.18, 0.20, 0.25 as the second group (#2) to test reproducibility of the initial synthesis. Afterwards, we synthesized additional samples at $\delta$ = 0.25 and 0.30 (#3). An allied sample LaFeAsO$_{0.85}$ was prepared for a comparison in the same manner. A La rod (99.9%, Nilaco Co.) was powdered by scratching the surfaces in a globe box, and was employed as a starting material instead of the Tb source.

The sample quality was investigated by a powder X-ray diffraction (XRD) method in a commercial apparatus (Rigaku, RINT2200V/PC) using Cu-K$\alpha$ radiation. Crystal structure of the selected samples LaFeAsO$_{0.85}$ and TbFeAsO$_{0.85}$ (#1) were studied by a synchrotron X-ray diffraction (SXRD) method at ambient temperature/pressure in a large Debye-Scherrer camera at the BL15XU beam line of SPring-8 [16]. Incident beam was monochromatized at $\lambda$ = 0.65297 Å, and a sample capillary (Lindenmann glass made, outer diameter was 0.1 mm) was rotated during the intensity measurement to reduce the preferred-orientation effect. Oxygen content of the selected samples was measured by a gravimetric method.

Electrical resistivity ($\rho$) of the polycrystalline TbFeAsO$_{1-\delta}$ was measured by a four-point probe method with a gage current of 0.1 mA ($\delta$ = 0.10) and 1 mA (0.15, 0.18, 0.20, 0.25, 0.30) in a Quantum Design PPMS. Electrical contacts on the four locations along the rectangular parallelepiped sample were prepared from gold wires and silver paste. The measurements were conducted between 2 K and 300 K. The $\delta$ = 0.15 sample (#1) was selected to study a magnetic field dependence ($\leq$ 90 kOe) of $\rho$. Magnetic susceptibility ($\chi$) of the samples was measured in a Quantum Design MPMS. The sample each was cooled to 2 K without applying a magnetic field, and then the sample was slowly warmed up in a magnetic field of 10 Oe (zero-field cooling, ZFC), followed by cooling down to 2 K (field cooling, FC). Isothermal magnetization of a selected sample was measured at 2 K up to 200 Oe in the MPMS. Specific heat ($C_p$) of the samples was measured in the PPMS by a time-relaxation method.

Ac magnetic susceptibility $\chi' - i\chi''$ was measured on the polycrystalline TbFeAsO$_{0.85}$ (#1) in a laboratory-made magnetometer, which comprises the Hartshorn bridge and a temperature controller. Balance of the electrical bridge was carefully maintained during the measurements. A two-phase lock-in amplifier was used to detect in-phase ($\chi'$) and out-of-phase ($\chi''$) signals of the bridge over the superconducting transition. An ac magnetic field was applied to the sample during the measurements without magnetic shielding; the applied ac field was thus superimposed on the Earth's field of ~300 mOe. We used $H_{ac}$ = 100 mOe, where a fundamental frequency was 132 Hz. Details of the magnetometer were reported elsewhere [17].

**Results and discussion**

Fig. 1 shows XRD profiles of the polycrystalline TbFeAsO$_{1-\delta}$ at $\delta$ = 0, 0.10, 0.15, 0.20, 0.25 (#1) and $\delta$ = 0.18 (#2) and $\delta$ = 0.30 (#3). We carefully compared the XRD profiles of the samples having the same $\delta$, and found no significant difference, suggesting high degree of synthesis reproducibility. Over the entire profiles, major peaks were clearly indexed by the *hkl* numbers as shown in ref. 15, calculated from a tetragonal structure model, while the $\delta = 0$ profile has non-trivial impurities peaks (marked by stars). It is notable that the low sample quality at $\delta = 0$ was dramatically improved by introducing an amount of the oxygen vacancy. We found even 10 mole % of the oxygen vacancy, i.e. $\delta = 0.10$, sufficiently improved the sample quality, suggesting two possibilities at least. (i) The nominal oxygen quantity was underestimated by approximately 0.1 mole per the formula unit (f.u.) and the oxygen-stoichiometry was practically realized at $\delta = 0.1$. (ii) High-pressure chemical reaction is sluggish in the oxygen-defect free system. A useful hint was obtained by subsequent magnetic studies (the data are shown later). A superconducting transition was observed at $\delta = 0.1$ and was not at all (>2 K) at $\delta = 0$. Considering the lattice parameters which were certainly changed by introducing the oxygen vacancies (Fig. 2), it is thus likely that the later is primarily responsible for the sample quality issue at $\delta = 0$.

In Fig. 2, we plotted the lattice parameters of the #1-3 samples simultaneously, confirming high reproducibility of the sample quality. The high reproducibility however does not secure that the nominal quantity $\delta$ is equivalent to the true quantity; we thus measured the net oxygen quantity as follows. We measured the weight of the high-pressure specimen (sample plus Pt tube) before and after the high-pressure run finding no substantial difference. Thus, the net oxygen content of the sample was kept unchanged during the heating. In order to confirm this we analyzed the net oxygen contents of selected samples by a gravimetric method. Nearly 10 mg of a selected sample was fully oxidized at 1450~1500 °C in air or oxygen then was slowly cooled. The final product was found to be $TbFeO_3$ by a XRD measurement indicating that the reaction of this procedure is $TbFeAsO_y$ + $(4.5-y)/2 O_2$ = $TbFeO_3$ + $1/2\ As_2O_3\uparrow$. From the weight loss during the procedure, the net oxygen contents were calculated to be 0.71 and 0.64 for the two selected samples with the nominal oxygen contents of 0.75 ($\delta = 0.25$) and 0.70 ($\delta = 0.3$), respectively. The experimental oxygen content is slightly smaller than the initial value but the agreement was fairly good.

Let us go back to see the $\delta$ dependence of the tetragonal cell dimension of $TbFeAsO_{1-\delta}$ shown in Figs 2a-2c. The unit-cell volume decreases with increasing δ and shortly hits the bottom at δ of 0.15-0.18: the change reaches -1.1 % almost isotropically (0.3 % shrinks along a-axis and 0.4 % along c-axis). The unit-cell volume then turns to increase slightly by further introducing amount of the oxygen-vacancies. The overall feature is thus out of the Vegard's law, resulting from a probable competition between the increased mean ionic size of Fe and the unit-cell size reduction due to the oxygen missing. Interestingly, alike feature including the broad minimum in the plots was observed for $NdFeAsO_{1-\delta}$ [18]. In large contrast, 0.2 % increment in the unit-cell volume was observed for the F doped Tb system [19].

We examined a relation between the lattice parameters and $T_c$: $T_c$ vs. $a$ is plotted in Fig. 2d. The $T_c$ was estimated from the magnetic data shown later. It is clear that the lattice parameter change is correlated with $T_c$ change, as found in $NdFeAsO_{1-\delta}$ and NdFeAs(O,F) [14,20]. It is notable that not only the 'under-doped' superconductor, but also the 'over-doped' superconductor likely follows a unique curve in contrast to what was observed for $NdFeAsO_{1-\delta}$ [21].

Figs. 3a-3e show temperature and δ dependence of the dc magnetic susceptibility of $TbFeAsO_{1-\delta}$ measured in a field of 10 Oe. At δ = 0, the sample does not show a superconducting drop down to 2 K, while others (δ = 0.1-0.30) show clearly superconducting transitions. The series of the #2 samples show nearly identical properties (shown by solid curves in red) with those for the same δ #1 samples, confirming reproducibility of the magnetic properties, except at δ = 0.25. A rigid shift of the curves along the vertical axis was observed for the δ = 0.25 #2 sample probably due to magnetic impurities in the sample accidentally produced.

The highest $T_c$ of 44 K was observed at δ = 0.18, and the shielding fraction was also optimized at δ = 0.18: it is 1.15 (5 K, ZFC curve), while 1.00 is expected for the prefect diamagnetism. Although superconducting transitions were observed between δ of 0.1 and 0.30, the transitions were relatively broad even at the $T_c$ optimized δ of 0.18. Regarding the copper oxide superconductors, a much sharp transition was commonly observed even for a polycrystalline sample. The broad transition thus suggests that further improvements of the sample quality might be possible.

The observed $T_c$ is plotted against $\delta$ in Fig. 3g, showing a nearly bell-shaped feature. Compared with the results for NdFeAsO$_{1-\delta}$ [13], the highest $T_c$ of 51 K was found at $\delta \sim 0.17$ by a neutron diffraction method, being comparable with the result for TbFeAsO$_{1-\delta}$. Although an "over-doped" feature was expected at $\delta > 0.17$ for NdFeAsO$_{1-\delta}$, the synthesis condition might not allow synthesizing the over-doped Nd sample, unfortunately.

In order to evaluate the lower critical field ($H_{c1}$), we measured the initial diamagnetic curve of the highest $T_c$ sample ($\delta = 0.18$) at 2 K; the curve is shown in Fig. 3h, indicating that $H_{c1}$ is approximately 20 Oe, being much lower than that of the other $Ln$FeAsO superconductors [13]. Meanwhile, it appears that the $H_{c1}$ is nearly comparable with the applied magnetic field in the magnetic susceptibility measurements, implying a possible origin in part of the relatively broad transitions.

Figs. 4a and 4b show temperature and $\delta$ dependence of $\rho$ of the polycrystalline TbFeAsO$_{1-\delta}$, showing varieties of $T_c$ and normal state resistivity. The $T_c$ on-set changes in a comparable way with what was found in the magnetic study. The maximum $T_c$ on-set ($\delta = 0.18$) estimated from the resistivity data by a graphical analysis is 44 K, being consistent with the magnetic bulk $T_c$. Meanwhile, the normal state behavior is rather unusual: for example, the room-temperature $\rho$ goes down with introducing the oxygen vacancies and hits the bottom at the highest $T_c$ composition. The room-temperature $\rho$ turns to increase by further accommodation of the oxygen vacancies. As it is widely accepted, charge transport in normal state is highly sensitive to the polycrystalline issue, including degree of sintering, grain boundaries, and surface impurities. Those factors often mask true nature of the charge transport. Thus, quantitative analysis regarding normal state conductivity is thus left for future work after a high-quality single crystal becomes available.

Fig. 5 shows temperature and $\delta$ dependence of $C_p$ of the polycrystalline TbFeAsO$_{1-\delta}$. First, lattice contribution was analyzed by a linear combination of the Debye model and the Einstein model,

$$C(T) = n_{\rm D} \times 9N_{\rm A}k_{\rm B}\left(\frac{T}{T_{\rm D}}\right)^3 \int_0^{T_{\rm D}/T} \frac{x^4 e^x}{(e^x-1)^2}dx + n_{\rm E} \times 3N_{\rm A}k_{\rm B}\left(\frac{T_{\rm E}}{T}\right)^2 \frac{e^{T_{\rm E}/T}}{\left(e^{T_{\rm E}/T}-1\right)^2},$$

where $N_{\rm A}$ is the Avogadro's constant, $k_{\rm B}$ is the Boltzmann's constant, and $T_{\rm D}$ and $T_{\rm E}$ are the Debye and the Einstein temperatures, respectively. The scale factors $n_{\rm D}$ and $n_{\rm E}$ correspond to the number of vibrating modes per the formula unit in the Debye and the Einstein models, respectively. The fit was carried out between 25 K and 80 K for every samples, and the best fit was obtained at $T_{\rm D}$ of 324(3) K, $T_{\rm E}$ = 80.6(5) K, $n_{\rm D}$ = 0.716(4)×3.90, $n_{\rm E}$ = 0.118(5)×3.90 for the δ = 0.10 sample, and $T_{\rm D}$ of 367(3) K, $T_{\rm E}$ = 89.0(7) K, $n_{\rm D}$ = 0.774(8)×3.70, $n_{\rm E}$ = 0.191(5)×3.70 for the δ = 0.30 sample. For the other samples between δ of 0.10 and 0.30, the refined parameter each was found to be intermediate between those, indicating the lattice contribution in $C_{\rm p}$ is less dependent on the oxygen quantity change. We found $T_{\rm D}$ for TbFeAsO$_{1-\delta}$ is almost comparable with $T_{\rm D}$ for the other superconducting iron compounds [22-24].

The fit to the $C_{\rm p}$ data (fat solid curves in Fig. 6) makes clear that the magnetic contribution at low temperature is substantially large. The sizable contribution probably comes from the localized magnetic moments of Tb$^{3+}$, which may be less relevant to the observed superconductivity. Detailed analysis of the large magnetic term is thus left for future work.

We carefully searched for an anomaly in the $C_{\rm p}$ data around $T_{\rm c}$, finding a step-like feature (indicated by arrows) at δ = 0.15, 0.18, and 0.20. For a convenience, temperature differential curve of the $C_{\rm p}$ is shown at the bottom of the panel each (thin solid curve), indicating a spike at $T_{\rm c}$ beyond statistical background. It appears that the step-like anomaly changes as a function of δ in a comparable way with what was observed in the magnetic and the electrical resistivity measurements. At δ = 0.1, 0.25, 0.30, expected anomalies are masked by the sizable Tb$^{3+}$ contribution.

Fig. 6 shows an expansion of the $C_{\rm p}/T$ vs. $T$ around $T_{\rm c}$ for the δ = 0.18 sample. The dotted curve is the lattice contribution estimated by the analysis using the Debye and the Einstein models as discussed above. The subtracted part is shown in the inset to Fig. 6, showing a superconducting jump.

The midpoint of the thermo-dynamical transition is 44 K, which matches well with the magnetic $T_c$. $\Delta C_p/T_c$ is estimated to be approximately 5 mJ mol$^{-1}$ K$^{-2}$. A simple BCS prediction indicates that the coefficient of the linear term in the specific heat in the normal state ($\gamma$) is $\Delta C_p/1.43T_c$ for a weakly coupled superconductor. From the present result, we thus estimated $\gamma$ of 3.5 mJ mol$^{-1}$ K$^{-2}$ for the $\delta$ = 0.18 sample. Due to the large Tb$^{3+}$ contribution and the relatively higher $T_c$, $\gamma$ is unable to be directly measured for a comparison. However, $\gamma$ of the related superconductors LaFe$_{1-x}$Co$_x$AsO (x = 0.05, 0.11, 0.15) was directly measured to be approximately 3-8 mJ mol$^{-1}$ K$^{-2}$ [23], being comparable with the present result for the Tb-1111 system. The F-doped LaFeAsO has also a comparable $\gamma$ of 3-8 mJ mol$^{-1}$ K$^{-2}$ [22]. The estimated $\gamma$ for the $\delta$ = 0.18 sample is thus not far from that for the other 1111 superconductors. Besides, the jump $\Delta C_p/T_c$ for the F-doped SmFeAsO was found approximately 6-8 mJ mol$^{-1}$ K$^{-2}$ [25], indicating $\gamma$ of the superconducting $Ln$FeAsO is much smaller regardless of size of $Ln$ than that of the hole-doped FeAs-122.

Obviously, the superconducting properties of TbFeAsO$_{1-\delta}$ are optimized around $\delta$ of 0.15-0.18. Fig. 7 shows ac susceptibility of the $\delta$ = 0.15 sample. The real part ($\chi'$) shows a steep change at $T_c$, which accords with what was observed in the dc susceptibility measurements. Besides, the imaginary part ($\chi''$) consists of a rather single peak than multiple peaks, suggesting high degree of chemical homogeneity of the polycrystalline sample. The inset to Fig. 7 shows magnetic field dependence of the electrical resistivity of the $\delta$ = 0.15 sample. At $H$ = 0 kOe, the superconducting transition is sharp, even though the sample is polycrystalline. The transition becomes slightly broad with increasing magnitude of the applied magnetic-field. A graphical analysis on the curves indicates that the $T_c$ on-set goes down ~2 K in a field of 90 kOe, suggesting high magnitude of the upper critical field ($H_{c2}$). Unfortunately, the experimental range is too narrow to properly estimate $H_{c2}(T = 0 K)$, however an attempt using an analytical formula $\mu_0 H_{c2} = -0.693 (dH_{c2}/dT)_{T=T_c} T_c$, developed by Werthamer, Helfand, and Hohenberg [26], indicated that the $\mu_0 H_{c2}(0)$ is near the Pauli-limit of 77.5 T ($\mu_0 H_{Pauli} = 1.24 k_B T_c / \mu_B$, where $T_c$= 42 K). It is thus clear at least that $H_{c2}(0)$ of TbFeAsO$_{0.85}$ is much lower than what was found for a single crystal NdFeAsO$_{0.88}$F$_{0.12}$ (300 T for $\mu_0 H_{c2,ab}$) [11].

Consequently, we studied the atomic structure of the $T_c$ optimized sample ($\delta$ = 0.15) by a SXRD method. The allied sample LaFeAsO$_{0.85}$ prepared under the same synthesis condition was studied for a comparison. The program RIETAN-2000 was used to analyze the SXRD patterns by means of a Rietveld method [27]. A neutron diffraction study for the F-doped LaFeAsO reported elsewhere suggested that the room-temperature structure is well described by a tetragonal model having symmetry *P*4/*nmm*, while another model having *Cmma* (orthorhombic) was found below 155 K [28]. We employed the tetragonal model as an initial model to parameterize the room-temperature structure of the oxygen vacant LaFeAsO$_{0.85}$ and TbFeAsO$_{0.85}$. Coefficients for analytical approximation to the atomic scattering factors were taken from ref. 29. The pseudo-Voigt function of Toraya was used as a profile function [30]. The SXRD background was characterized by an 11th-order Legendre polynomial function. Isotropic atomic displacement parameters ($B_{iso}$) and isotropic Debye-Waller factor, $\exp(-B_{iso}\sin^2\theta/\lambda^2)$, were assigned to all the atoms.

First, we studied the structure of LaFeAsO$_{0.85}$. In a course of the SXRD profile analysis, small contributions from impurities, presumably Fe$_2$O$_3$ and FeAs, were detected. Since the impurity peaks were too small to estimate those crystallographic parameters correctly, we refined only those scale factors and lattice constants. Final refinements for LaFeAsO$_{0.85}$ were conducted simultaneously with the impurities peaks. The weight fraction finally estimated was 6.7 % for Fe$_2$O$_3$, 0.81% for FeAs, and 92.5% for LaFeAsO$_{0.85}$. The *R* factors in total were below 2.3 % and the difference curve was found fairly smooth (shown at the bottom of Fig. 8a), indicting high-quality of the solution. It appeared that the reasonable solution for LaFeAsO$_{0.85}$ was indeed attained by using the tetragonal model. At the end of the analysis, we fixed the oxygen thermal parameter to test stability of the oxygen occupancy factor in the analysis. A brief summary of the SXRD analysis is in Table I. The result secured that the tetragonal structure model and the analysis method appropriate for the oxygen vacant system prepared under the high-pressure conditions.

We analyzed the SXRD profile for the δ = 0.15 sample as well. We repeated the same steps applied for LaFeAsO$_{0.85}$; the refinement steadily reached a reasonable solution. The *R* factors were below 2.9 %. The impurities Fe$_2$O$_3$ and FeAs were not detected above the background, while small amount of other impurities including TbAs [15] were found instead. Although the *R* factors were certainly low, the deference curve especially in the vicinity of the main peak (Fig. 8b) suggests further improvement is likely possible. Additional studies including an electron microscopy of the oxygen vacancies and local structure distortion/tilting may assist further improvements.

Insets to Figs. 8a and 8b show structure images, drawn on the present results for the δ = 0.15 sample. Since the coordination environment of Fe was suggested to be coupled with the superconducting properties [14], we carefully investigated the bond angle between Fe and coordinated As for both LaFeAsO$_{0.85}$ and TbFeAsO$_{0.85}$. The calculated bond angles (As-Fe-As) were 113°, 108° for LaFeAsO$_{0.85}$, and 109°, 110° for TbFeAsO$_{0.85}$. It appeared that the angles are fairly close to 109.47° for the regular tetrahedron. As discussed in ref 14, $T_c$ is probably optimized at the ideal angle. It is thus interesting to investigate possible correlation between the bond angle and amount of oxygen vacancies, which may be coupled with $T_c$. In order to evaluate the possible correlations further, additional studies are in progress.

In the F-doped *Ln*FeAsO system, a possible accompany of oxygen vacancies with the dopant is usually unexcused and its role remains ambiguous; in short, the true chemical composition should be much closer to *Ln*FeAsO$_{1-δ}$F$_x$ than *Ln*FeAsO$_{1-x}$F$_x$ [31]. Studies of the F-free superconductors *Ln*FeAsO$_{1-δ}$ are thus expected to exclude additional complexity. In this study, a nearly bell-shaped feature in $T_c$ vs. δ (summary is shown in Fig. 9) was observed for the first time to our best knowledge possibly because the complexity is excluded to some extent. Perhaps, the smaller element of Tb may also be help to accept such the relatively large amount of oxygen vacancies in the structure, tuning the supercomputing state from the "under-doped" to the "over-doped". The chemical analysis confirmed the true oxygen quantity is close to the nominal.

At the end, the similarity of the bell-shaped feature between the iron and the copper oxide superconductors implies possible common physics. We have thus an open question that what is the principal origin to reduce the $T_c$ in the "over-doped" region. Since an "over-doped" superconductivity is found for the Fe superconductor as well as for the Cu superconductor, continuous efforts are directed to investigate nature of the superconductivity of the "over-doped" TbFeAsO$_{1-\delta}$ [32].

**Acknowledgments**

This research was supported in part by World Premier International Research Center (WPI) Initiative on Materials Nanoarchitectonics from MEXT, Japan, Grants-in-Aid for Scientific Research (20360012) from JSPS, Japan.

Table I. Structure parameters of LaFeAsO$_{0.85}$ and TbFeAsO$_{0.85}$ determined by a synchrotron X-ray powder diffraction method at room temperature. Space group is *P*4/*nmm* (no. 129), *Z* = 2, *a* = 4.02397(4) Å, *c* = 8.71513(8) Å, *V* = 141.118(2) Å$^3$, and $\rho_{cal}$ = 6.664 g/cm$^3$ for LaFeAsO$_{0.85}$, and *a* = 3.89320(4) Å, *c* = 8.38458(14) Å, *V* = 127.085(3) Å$^3$, and $\rho_{cal}$ = 7.887 g/cm$^3$ for TbaFeAsO$_{0.85}$. *R* factors were $R_{wp}$ = 2.15 %, $R_p$ = 1.48 %, and $R_F$ = 2.29 % for LaFeAsO$_{0.85}$, and $R_{wp}$ = 2.83 %, $R_p$ = 2.13 %, and $R_F$ = 1.88 % for TbaFeAsO$_{0.85}$.

| Site | Wyckoff position | g | x | y | z | B (Å$^2$) |
|---|---|---|---|---|---|---|
| La | 2c | 1 | 0.25 | 0.25 | 0.14469(5) | 0.557(8) |
| Fe | 2b | 1 | 0.75 | 0.25 | 0.5 | 0.526(18) |
| As | 2c | 1 | 0.25 | 0.25 | 0.65367(8) | 0.692(13) |
| O | 2a | 0.845(5) | 0.75 | 0.25 | 0 | 1.5 * |
| Tb | 2c | 1 | 0.25 | 0.25 | 0.13766(6) | 0.709(12) |
| Fe | 2b | 1 | 0.75 | 0.25 | 0.5 | 0.99(3) |
| As | 2c | 1 | 0.25 | 0.25 | 0.66502(11) | 0.94(2) |
| O | 2a | 0.76(1) | 0.75 | 0.25 | 0 | 1.5 * |

* Fixed values.

Selected bond lengths (*d*) and angles (θ) are $d_{Fe-As}$ = 2.4170 Å (×4), $d_{La-O}$ = 2.3745 Å (×4), $\theta_{Fe\text{-}As\text{-}Fe}$ = 112.701°, 72.119°, $\theta_{As\text{-}Fe\text{-}As}$ = 107.881, 112.702° for LaFeAsO$_{0.85}$, and $d_{Fe-As}$ = 2.3883 Å (×4), $d_{Tb-O}$ = 2.2631 Å (×4), $\theta_{Fe\text{-}As\text{-}Fe}$ = 109.190°, 70.388°, $\theta_{As\text{-}Fe\text{-}As}$ = 109.612, 109.190° for TbFeAsO$_{0.85}$.

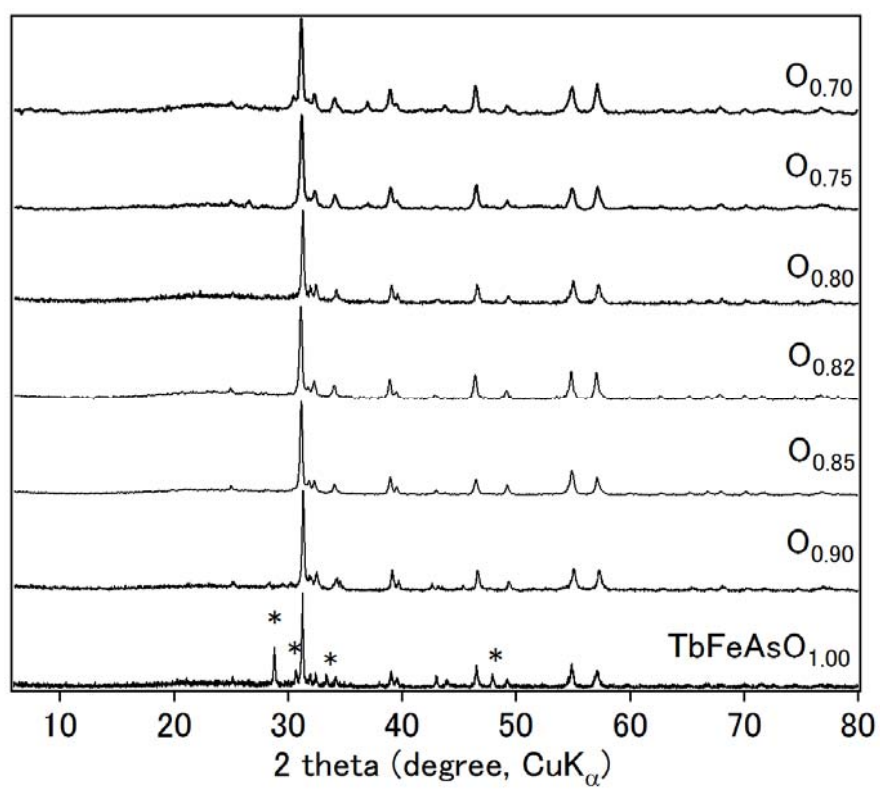

Fig. 1   XRD profiles of the polycrystalline TbFeAsO$_{1-\delta}$.  Nominal composition is indicated at the profile each.   Peak identification was successful with the tetragonal structure model (*P*4/*nmm*) as was in ref. 15.   Stars indicate peaks due to impurities including Tb$_2$O$_3$.

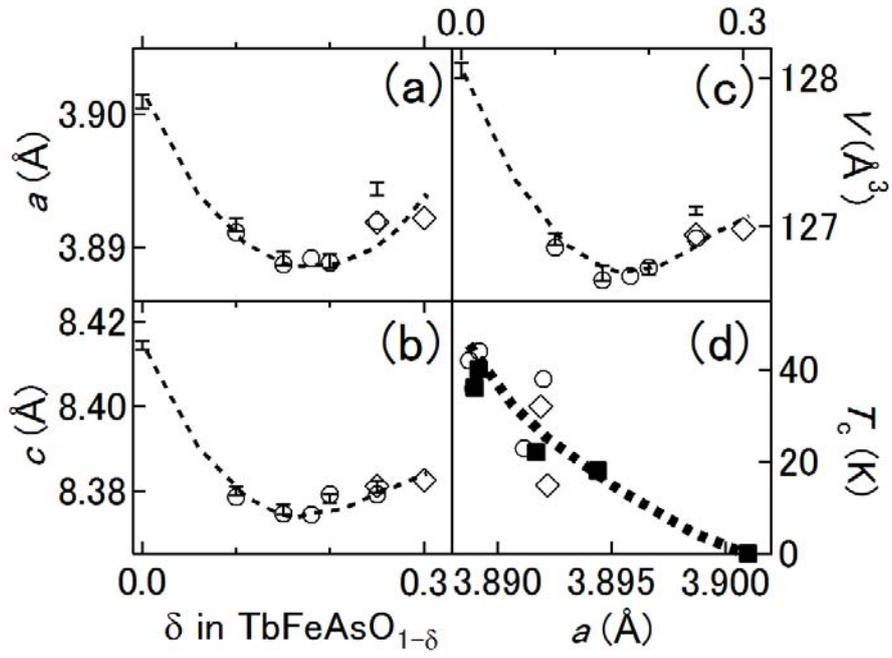

Fig. 2  Unit-cell size evolution of TbFeAsO$_{1-\delta}$ through the oxygen quantity; (a-c) show *a*, *c*, and the cell volume, respectively.  (d) shows magnetic $T_c$ vs. *a*.  Open symbols are data obtained from the #2 (circle) and the #3 (diamond) groups of the samples.  Dotted curves are guide to the eye.

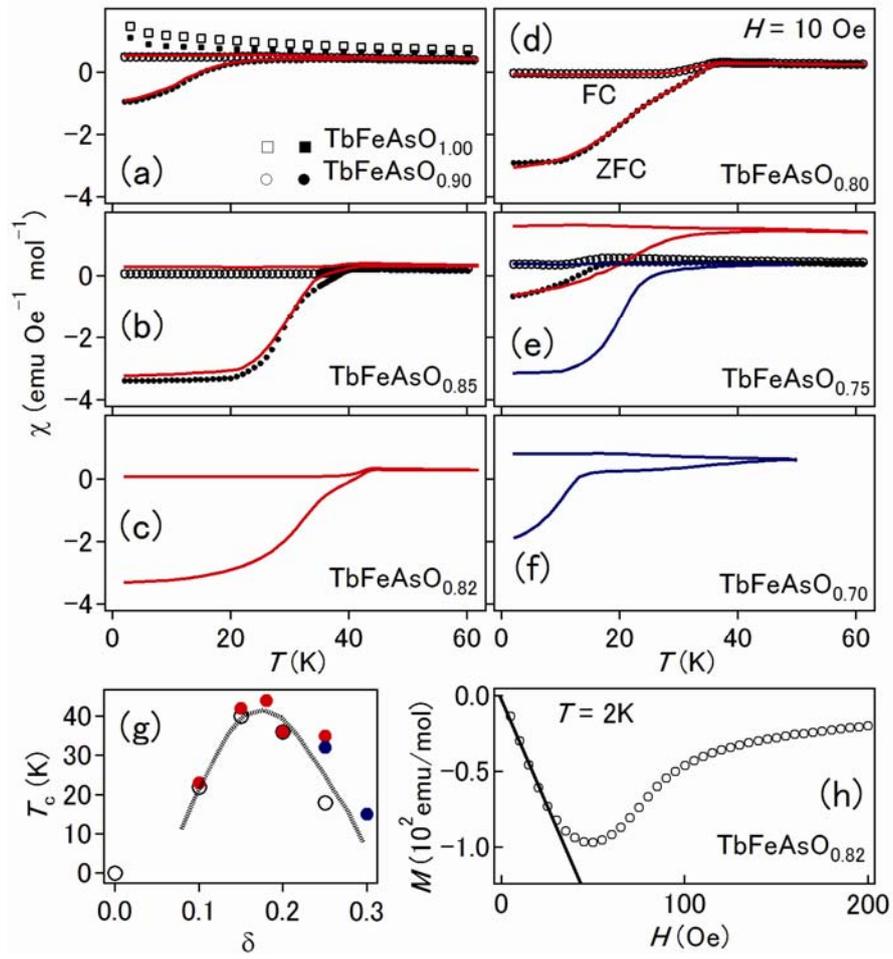

Fig. 3 Temperature and oxygen concentration dependence of the dc magnetic susceptibility ($H$ = 10 Oe) of the polycrystalline TbFeAsO$_{1-\delta}$, where (a) $\delta$ = 0, 0.1, (b) 0.15, (c) 0.18, (d) 0.20, (e) 0.25, and (f) 0.30. The solid curves in red and blue are data for the #2 and #3 groups of the samples. (g) Summary of the magnetic $T_c$ vs. the oxygen concentration. (h) shows the initial isothermal magnetization (2 K) of the $T_c$ optimized sample TbFeAsO$_{0.82}$.

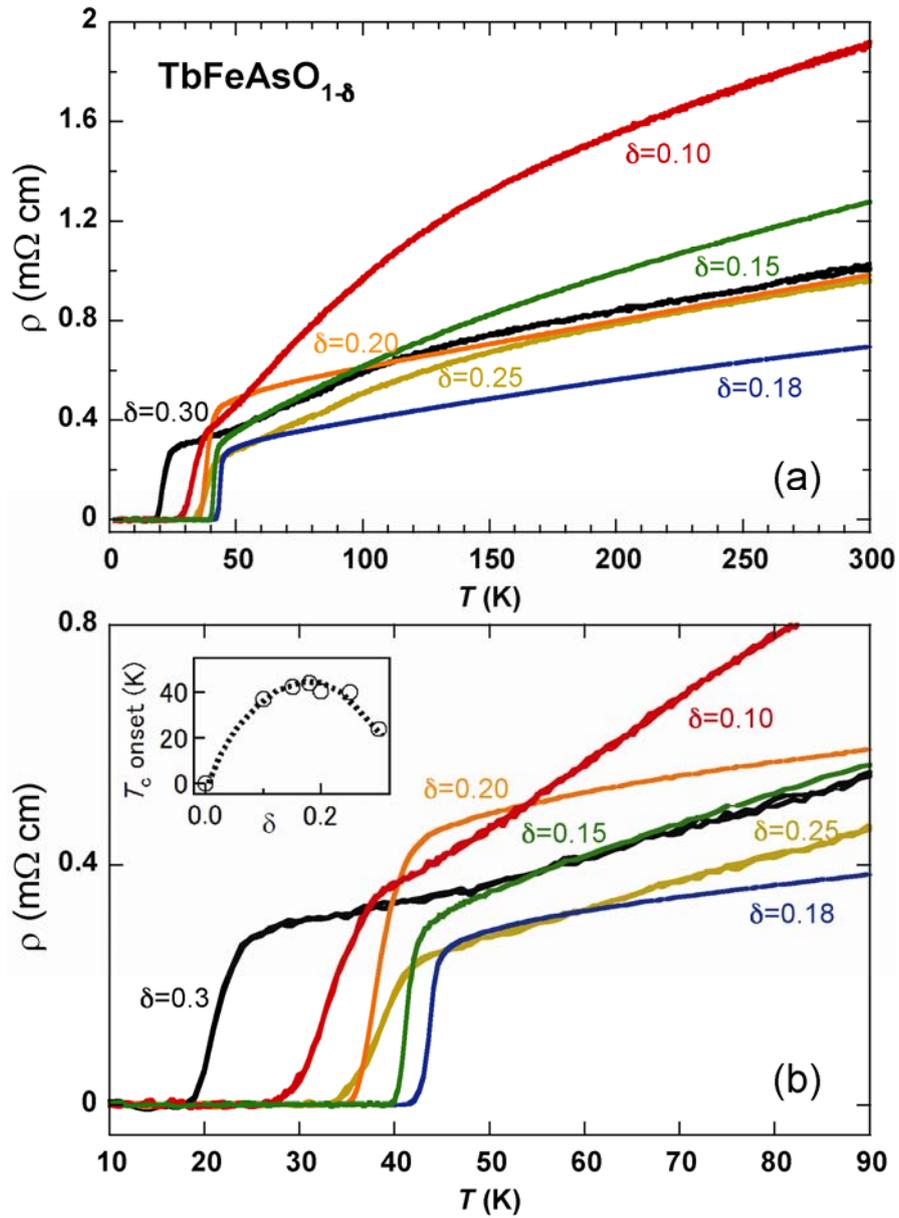

Fig. 4　(a) Temperature and oxygen concentration dependence of the electrical resistivity of the polycrystalline TbFeAsO$_{1-\delta}$, and (b) the expansion of the data.　Inset shows $T_c$ onset vs. $\delta$.

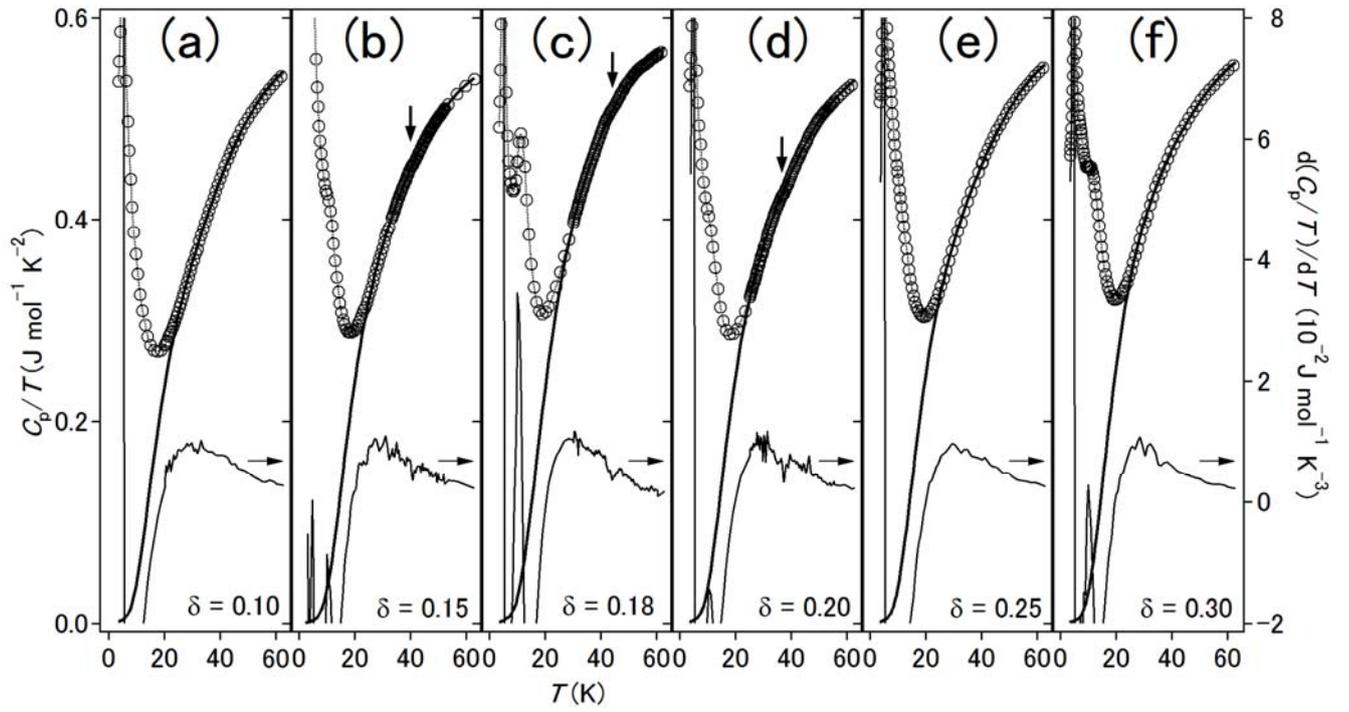

Fig. 5 Oxygen concentration and temperature dependence of the specific heat of the polycrystalline TbFeAsO$_{1-\delta}$, where (a) $\delta$ = 0.1, (b) 0.15, (c) 0.18, (d) 0.2, (e) 0.25, and (f) 0.30. Fat solid curves are fit to the data by using the Debye and Einstein lattice models, and thin solid curves are temperature differential of the data.

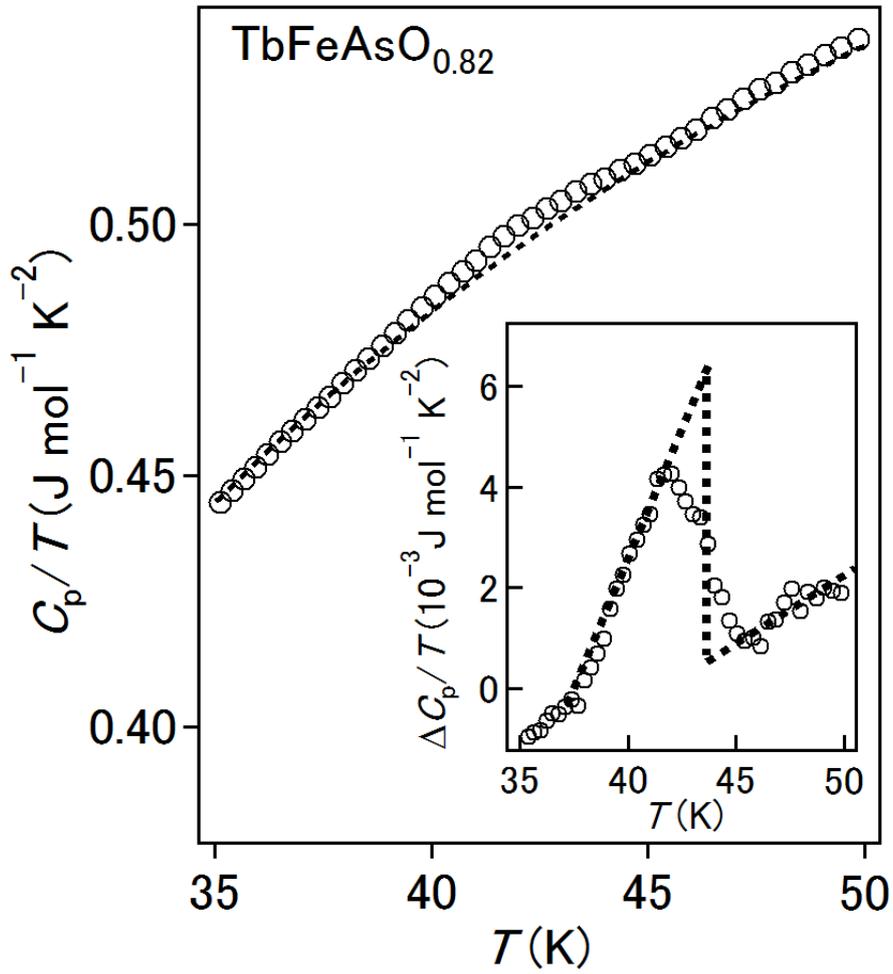

Fig. 6　Expansion of the $C_p$ data for the $T_c$ optimized sample.　The fit to the data by the lattice model was shown as a dotted curve and the difference between the data and the fit is shown in the inset.

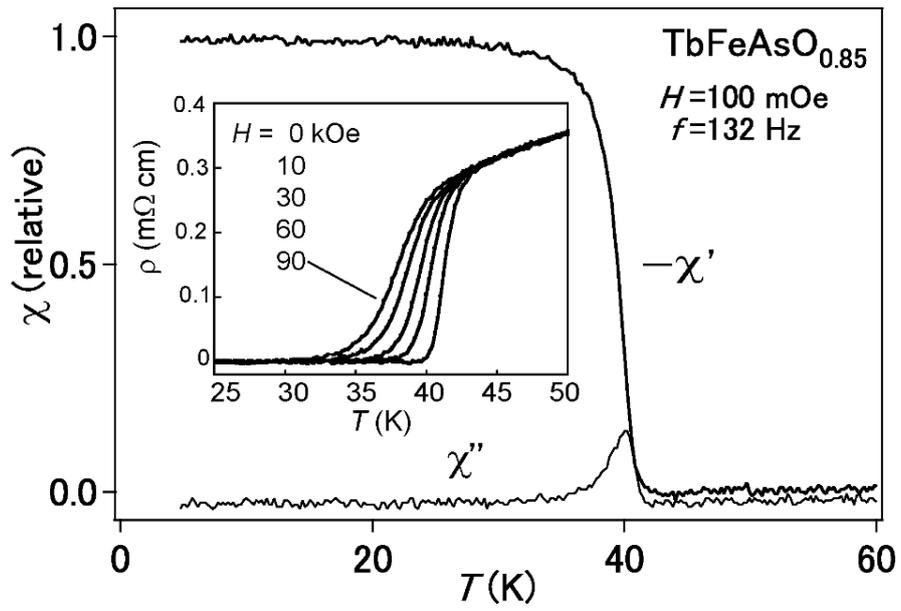

Fig. 7  Temperature dependence of the ac magnetic susceptibility and (inset) temperature and magnetic field dependence of the electrical resistivity of the polycrystalline TbFeAsO$_{0.85}$.

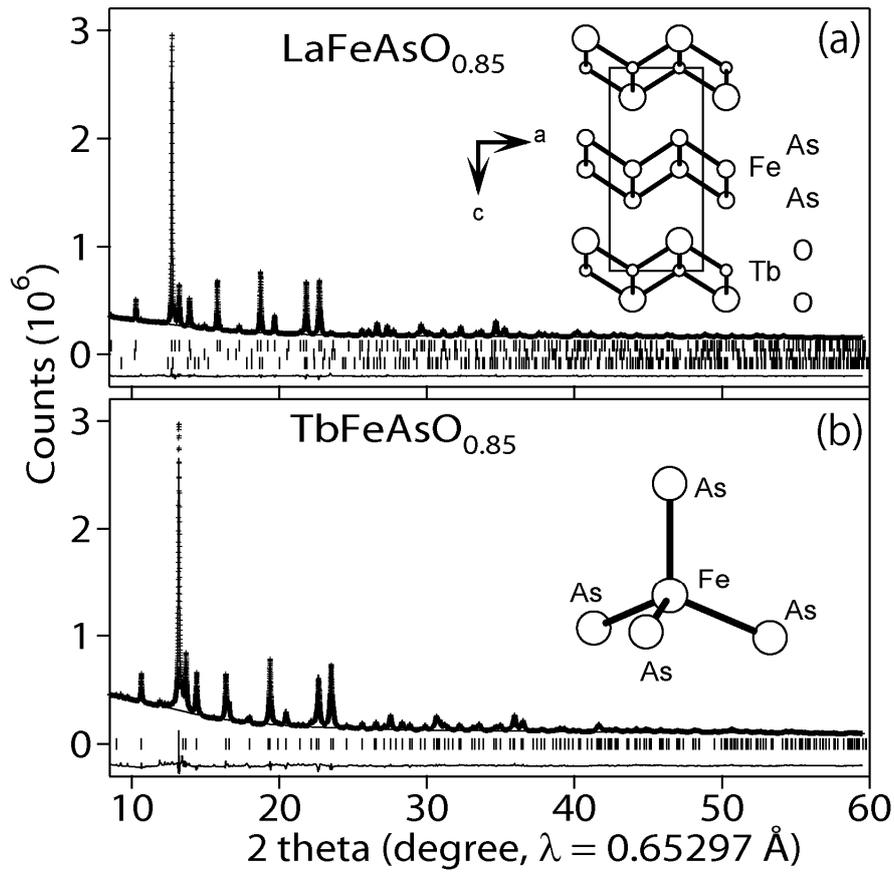

Fig. 8   Rietveld analysis of synchrotron X-ray profiles of (a) LaFeAsO$_{0.85}$ and (b) TbFeAsO$_{0.85}$. Cross markers and solid lines show the observed and calculated profiles, respectively.   The difference curve is at the bottom each.   The positions of Bragg reflections are marked by ticks.   Center and bottom lines of ticks are for the impurities Fe$_2$O$_3$ and FeAs, respectively.   Insets show structural images, based on the present result.

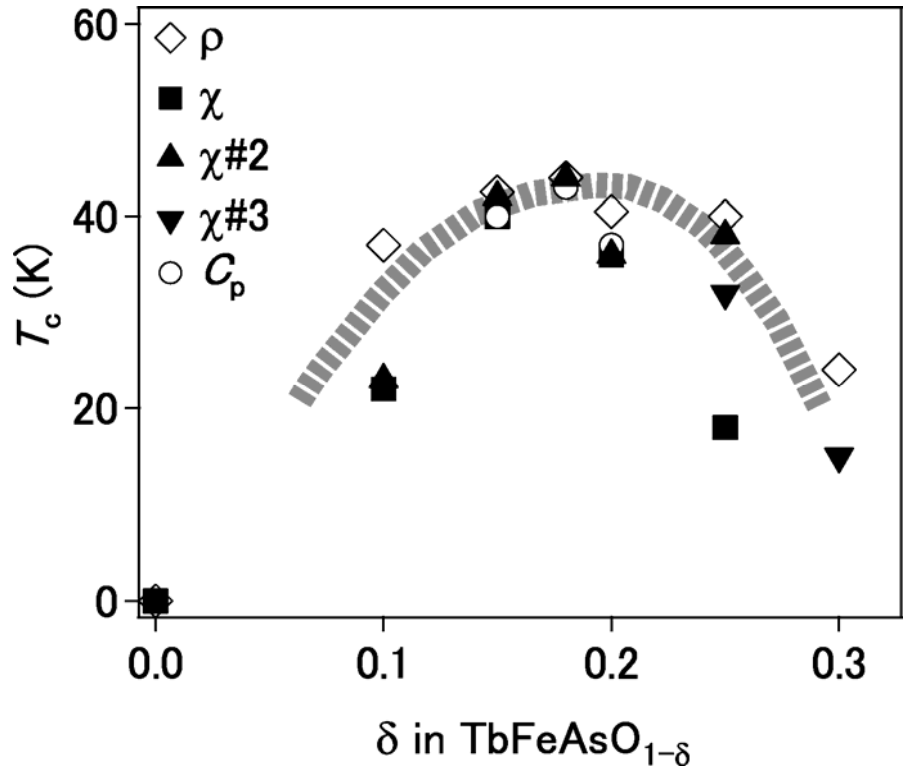

Fig. 9  Summary of $T_c$ change observed for TbFeAsO$_{1-\delta}$ through varieties of the properties measurements.  The marks #2 and #3 indicate data points for the #2 and #3 groups of the samples, respectively.